# On Thermal Vacuum Radiation of Nanoparticles and their Ensembles

G. V. Dedkov and A. A. Kyasov

Nanoscale Physics Group,  Kabardino-Balkarian State University, Nalchik, 360004, Russia

E-mail: gv_dedkov@mail.ru

Radiant emittance of dimers and ensembles of particles consisting of gold, graphite and silica glass nanoparticles in vacuum is studied numerically based on the fluctuation-electromagnetic theory. The presence of neighboring particles of the same temperature causes an oscillating character of radiant emittance (per one particle) depending on the particle size, interparticle distance and temperature. We conclude that an ensemble of particles could be a much more intense source of thermal radiation than an equivalent solid body with the same outer surface area. Alternatively, when the neighboring particles create a significant "screening effect" (silica), an ensemble of particles could be a very good heat protector.
PACS: 44.40+a; 65.80.+n; 65.90.+i.

## Introduction

A role of paramount importance in nanoelectronics, nanophotonics, and photovoltaics is played by heat dissipation.  Radiative heat exchange processes provide a convenient way of handling the excess of heat produced in these devices [1]. On the other hand, there is a great interest in studying the properties of high-temperature clusters [2] and dust particles in fusion devices [3] where the processes of radiative heat transfer between the particles and the surrounding vacuum are essential in the energy balance. However, in describing thermal vacuum radiation of small particles of submicron dimensions the Planck's law of blackbody radiation is not sufficient.

   As is well known, blackbody radiation mediates heat exchange between bodies placed in vacuum and separated by large distance $d$ compared to the thermal wavelength $\lambda_T = 2\pi\hbar c / k_B T$. For the two parallel plates of area $S$ that leads to the radiative heat transfer rate (RHT) corresponding to the Stefan-Boltzmann law $W_{SB} = \frac{\pi^2}{60} \frac{k_B^4}{\hbar^3 c^2} S T^4$  independent of $d$. However, at $d < \lambda_T$ the rate is enhanced greatly due to the involvement of evanescent waves [4,5] and satisfies a $1/d^2$ dependence at small distances $d$, as show the experiments [6,7].

In the case of two spherical particles of radius $r$ separated by a small vacuum gap of width $d$, a leading near-field contribution to RHT obeys the law $1/(2r+d)^6$ [8—10]. In [11], the authors have reconsidered this problem within the assumption of dipolar response by taking into account radiative corrections and electromagnetic crossed terms mixing the electric and magnetic particle responses. In particular, it was shown that crossed terms could play a dominant role in heterogeneous structures, and become less important in homogeneous ones when calculating RHT between the two particles of different temperature.

In this regard, no attention to date was paid to thermal radiation emitted into the surrounding vacuum provided that neighboring particles have the same temperature. Since the near-field interactions can be significant, the presence of another particle (or an ensemble of particles) affects thermal radiation of each particle when the ensemble and vacuum environment are characterized by different temperatures. An intriguing question is whether might the energy radiated by a particle be higher than the corresponding black-body limit or not, and how the answer depends on material properties, the size and spatial configuration of particles. It is worth noting that the finite-size effect of thermal radiation can be traced even within the simplest quantum approach, since there exists a non-zero minimum for the radiation frequency due to the finite size of the body [12]. Thermal radiation of nanoparticles was also considered in [13] using a simplified model approach.

In this work, based on the fluctuation-electromagnetic theory, we calculate RHT ($W$) for a spherical particle of radius $r$ and temperature $T_1$ which is a part of dimer or an ensemble of particles at temperature $T_1$ embedded in the environment at temperature $T_2$. We use the assumption of dipolar response within the Mie theory [14] and tabulated dielectric characteristics of materials from [15]. Despite that net RHT between the particles is absent, each of them emits (absorbs) thermal photons into (from) vacuum background and the rate of this process depends on a spatial configuration of particles embedded in vacuum. In the simplest case of two particles (dimer) separated by a center-to-center distance $R \geq 2r$, the cooperative effect and RHT between the particles and vacuum environment prove to be the functions of $R$.

Numerical calculations were performed for gold, graphite and silica glass (SiO$_2$) particles with radii ranging from 0.1 to $1 \div 3\,\mu m$ in the temperature range from 100K up to the melting temperature of the corresponding bulk materials. We calculated a normalized RHT (blackness factor) $W/W_{SB}$ depending on the particle radius and temperature. The sharp difference in the dependence of the blackness factor on temperature and radius was found for the particles of gold and graphite on the one hand, and SiO$_2$ particles on the other hand. In the former case, the maximum ratio of $W/W_{SB}$ is observed for particles with a radius of $0.15 \div 0.35\,\mu m$, increasing with temperature. For graphite particles, $W/W_{SB}$ reaches 0.92 at $T = 3000K$, but for gold particles the highest blackness factor does not exceed 0.084 at $T = 1337K$. Interestingly, the blackness factor for SiO$_2$ particles reaches the maximum value of 0.55 at a much greater particle radius of $3.4\,\mu m$ and a much lower temperature of $223K$.

In the case of dimers, the dependence of $W/W_{SB}$ on the distance $R$ between the particles demonstrates an oscillating character. The presence of a neighboring particle may lead either to a decrease of RHT ("screening effect"), or to an increase of RHT ("mirror effect"), depending on the particle radius and temperature. The highest screening (up to 9 %) is found for SiO$_2$ particles of radius $1\,\mu m$ at a temperature of $600K$, while the highest "mirror effect" does not exceed 3 % for all types of the particles.

For an ensemble of $N$ closely packed nanoparticles, their total RHT in vacuum rises proportionally to $N$, while the black body radiation from a solid body with the same outer radius rises as $N^{2/3}$. Therefore, at $N \gg 1$ the ensemble of nanoparticles should be a much more intense source (absorber) of thermal radiation than an equivalent solid body of the same size.

## Theory

Figure 1 shows the simplest system under study, namely the case of two identical isotropic spherical particles 1,2 in the vacuum environment. We assume that the particles are nonmagnetic and have the frequency-dependent electric and magnetic dipole polarizabilities $\alpha_{1,2}^{(e)}(\omega)$, $\alpha_{1,2}^{(m)}(\omega)$. Within the framework of fluctuation electrodynamics, the rate of heating (cooling) of each particle (particle 1 for definiteness) can be represented in the form [16]

$$\dot{Q} = \langle \dot{\mathbf{d}}_1(t)\mathbf{E}(\mathbf{r}_1,t)\rangle + \langle \dot{\mathbf{m}}_1(t)\mathbf{B}(\mathbf{r}_1,t)\rangle = \int_{-\infty}^{+\infty}\frac{d\omega'}{2\pi}\frac{d\omega}{2\pi}(-i\omega)\exp[-i(\omega+\omega')]\cdot$$
$$[\langle d_{1,i}(\omega)E_i(\mathbf{r}_1,\omega')\rangle + \langle m_{1,i}(\omega)B_i(\mathbf{r}_1,\omega')\rangle]$$
(1)

Here $\mathbf{d}_1(t),\mathbf{m}_1(t)$ – electric (magnetic) dipole moments of particle 1, $\mathbf{E}(\mathbf{r}_1,t),\mathbf{B}(\mathbf{r}_1,t)$ – spontaneous electric and magnetic field at the location point $\mathbf{r}_1$ of particle 1, $d_{1,i}(\omega), m_{1,i}(\omega), E_i(\mathbf{r}_1,\omega'), B_i(\mathbf{r}_1,\omega')$ are the corresponding projections of the Fourier-transforms, points over variables indicate the time derivatives. It should be noted that all the aforementioned quantities include spontaneous and induced components of the dipole moments and fields. More details are given in [17], where we have calculated the heat transfer and other important characteristics in the system of two rotating particles placed in vacuum. The basic formula which we need follows from Eq. (30) in [17] assuming that $\Omega = 0$ (rotation frequency) and making use the replacements $T_3 \to T_2, T_2 \to T_1$:

$$\dot{Q} = -\frac{2\hbar}{\pi c^3}\int_0^\infty d\omega \omega^4 \, \text{Im}\, \alpha_1^{(e)}(\omega)\left[\coth\frac{\hbar\omega}{2k_B T_1} - \coth\frac{\hbar\omega}{2k_B T_2}\right] - \frac{2\hbar}{\pi}\int_0^\infty d\omega\left(\frac{\omega^2}{\hbar c^2}\right)^2 \cdot$$
$$\cdot\left\{\text{Im}\,\alpha_1^{(e)}(\omega)\,\text{Im}\, D_{ik}(\omega,\mathbf{R})\,\text{Re}\left[\alpha_2^{(e)}(\omega)D_{ik}(\omega,\mathbf{R})\right]\cdot\left[\coth\frac{\hbar\omega}{2k_B T_1} - \coth\frac{\hbar\omega}{2k_B T_2}\right]\right\} +$$
$$+ (e \to m) + (e,m) \quad (2)$$

$$D_{ik}(\omega,\mathbf{R}) = \left(-\frac{\hbar c^2}{\omega^2}\right)\exp(i\omega R/c)\left[\left(\frac{\omega^2}{c^2 R} + \frac{i\omega}{cR^2} - \frac{1}{R^3}\right)(\delta_{ik} - n_i n_k) + 2\left(\frac{1}{R^3} - \frac{i\omega}{cR^2}\right)n_i n_k\right] \quad (3)$$

where $D_{ik}(\omega,\mathbf{R})-$ components of the retarded Green's function of photons in vacuum ($\mathbf{n} = \mathbf{R}/R$, $i,k = x,y,z$), the term $(e \to m)$ is identical to the first two terms and describes magnetic polarization effects, while the last term $(e,m)$ corresponds to the crossed magnetic-electric polarization contributions. As was shown in [11], the crossed terms can be significant in heterogeneous metal-dielectric systems. Here we consider only the case of metal-metal and dielectric-dielectric combinations neglecting these crossed terms. Moreover, Eq. (2) does not include the processes of multiple scattering, since they are also negligibly small [11]. If the particles have different temperature, Eq. (2) involves an extra term describing direct heat exchange between the particles [8—11]. In addition, it is worth noting that the first term in (2) corresponds to the limiting case $R \to \infty$ or if we neglect the cooperative effect of another particle [10], which is described by the second term of Eq. (2).

After substituting (3) in (2) the second term of (2) takes the form

$$\dot{Q}^{(2)} = -\frac{2\hbar}{\pi}\int_0^\infty d\omega\omega\,\text{Im}\,\alpha_1^{(e)}(\omega)\left[\text{Re}\,\alpha_2^{(e)}(\omega)f_1(\omega R/c) - \text{Im}\,\alpha_2^{(e)}(\omega)f_2(\omega R/c)\right]\cdot$$
$$\cdot\left[\coth\frac{\hbar\omega}{2k_B T_1} - \coth\frac{\hbar\omega}{2k_B T_2}\right] + (e \to m) \quad (4)$$

$$f_1(x) = 2(x^3 - 3x)\cos(2x) + (x^4 - 5x^2 + 3)\sin(2x) \quad (5)$$

$$f_2(x) = 2x^4 - 4x^2 + 6 - 2(x^4 - 5x^3 + 3)\cos(x)^2 + 2(x^3 - 3x)\sin(2x) \quad (6)$$

Therefore, radiant emittance $W$ of particle 1 is described by a sum of the first term in (2), involving direct RHT (referred to as $\dot{Q}^{(1)}$) between the particle and vacuum, and $\dot{Q}^{(2)}$ – indirect RHT due to the presence of the second particle (the second term in (2)). Assuming that $\dot{Q}^{(1)} < 0$, the sign of $\dot{Q}^{(2)}$ corresponds to the "mirror effect" ($\dot{Q}^{(2)} < 0$) or "screening effect" ($\dot{Q}^{(2)} > 0$) at a given $R$.

For an ensemble of $N$ equidistant closely packed particles of the same radius $r$ and temperature $T_1$, neglecting by the second and higher order correlations, as well as the surface effects, a total radiation power of the ensemble is given by

$$W_N \approx N\left(\dot{Q}^{(1)} + 12 \dot{Q}^{(2)}\right) \tag{7}$$

In numerical calculations, it is convenient to introduce the blackness parameter $b = W/W_{SB}$ or $b_N = W_N/W_{SB}$, relating to a single particle or to an ensemble of $N$ particles, where the black body radiant emittance $W_{SB}$ is given by

$$W_{SB} = \frac{\pi^3}{15} \frac{k_B^3}{\hbar^3 c^2} R_0^2 T^4 \tag{8}$$

with allowance for $R_0 = r$ in the former case and $R_0 = RN^{1/3}$, $R \geq 2r$ in the latter case.

## Numerical calculations

For graphite, we used numerical data [15] for the complex refractive index to obtain the dielectric function $\varepsilon(\omega)$ in the range of $0.01 \div 10\, eV$. These data correspond to the electric vector **E** perpendicular to the symmetry axis $c$ which is perpendicular to the basal plane of graphite. Tabulated data were approximated by cubic splines. The data for gold [15] were approximated by the Drude dielectric function $\varepsilon(\omega) = 1 - \omega_p^2/\omega(\omega + i\gamma)$ with $\omega_p = 9\, eV$, $\gamma = 0.053\, eV$ (at $T = 300K$). The material temperature dependence of gold was accounted for assuming that $\gamma$ increases linearly up to $0.214\, eV$ at $T = 1300K$. The data for SiO$_2$ [14] were taken in numerical form similarly to graphite and material parameters in these two cases were used without temperature corrections.

The scattering coefficients of electromagnetic field corresponding to the magnetic and electric polarization ($m, e$) of spherical particles within the Mie theory were calculated by the formulas [13]

$$t_n^{(m)} = \frac{-j_n(r_0)r_1 j_n'(r_1) + r_0 j_n'(r_0) j_n(r_1)}{h_n^{(+)}(r_0) r_1 j_n'(r_1) - r_0 \left[h_n^{(+)}(r_0)\right]' j_n(r_1)} \tag{9}$$

$$t_n^{(e)} = \frac{-j_n(r_0)\left[r_1 j_n(r_1)\right]' + \varepsilon(\omega)\left[r_0 j_n(r_0)\right]' j_n(r_1)}{h_n^{(+)}(r_0)\left[r_1 j_n(r_1)\right]' - \varepsilon(\omega)\left[r_0 h_n^{(+)}(r_0)\right]' j_n(r_1)} \tag{10}$$

where $r_0 = \omega r/c$, $r_1 = r_0 \sqrt{\varepsilon(\omega)}$, $\mathrm{Im}\, r_1 > 0$, $h_n^{(+)}(x) = n_n(x) + \mathrm{i} \cdot j_n(x)$, $j_n(x)$ and $n_n(x)$ are spherical Bessel functions, and the prime denotes differentiation with respect to $r_0$ and $r_1$. Dipolar polarizabilities $\alpha^{(e,m)}$ are obtained from (9), (10) at $n = 1$ via the relation $\alpha_n^{(q)} = (3c^3/2\omega^3) t_n^{(q)}$ ($q = e, m$). Following [11], we use the absorption factor $\beta^{(q)} = \mathrm{Im}\,\alpha^{(q)} - (2\omega^3/3c^3)|\alpha^{(q)}|^2$ in Eqs. (2) and (4) instead of $\mathrm{Im}\,\alpha^{(e,m)}$.

Numerical calculations of RHT were carried out assuming that $T_1 = T$ (hot particles) and $T_2 = 0$ (cold environment). Using Eq. (2), radiant emittance will then be defined as $W = -(\dot{Q}^{(1)} + \dot{Q}^{(2)})$. Figures 2—4 correspond to dimer particles of gold, graphite and SiO$_2$, correspondingly. In figures 2(a,b)—4(a,b), dependences of blackness parameter $b$ vs. particle radius $r$ and temperature $T$ are analyzed, assuming that $R/r = 2$ (close contact of dimer particles). Small irregularities on the curves are due to the "screening" ("mirror") effect of the second particle. In the case of gold and graphite particles, parameter $b$ reaches the maximum values of 0.084 and 0.92 at $T = 1337 K$ (melting temperature of gold) and $T = 3000 K$ respectively, while the corresponding radii are close to 0.16 and 0.35 $\mu m$ (Fig. 2a and Fig. 3a). In contrast to that, the blackness factor for SiO$_2$ particles reaches the maximum value of 0.55 at a much lower temperature of $223 K$ and a much higher radius of 3.4 $\mu m$ (Fig. 3a). Moreover, we can note quite different dependences of blackness factor $b$ on the particle radius $r$ and temperature $T$ for $Au$ and graphite particles on the one hand, and for $SiO_2$ particles on the other hand. In the former case, the maxima of $b$ are shifted to the side of lower $r$ with increasing $r$ (Figs. 2a, 2b), while for $SiO_2$ particles the maxima are shifted to higher $r$ with increasing $r$ (Fig. 4a). Dependences $b(T)$ demonstrate another behavior: for gold and graphite particles, the maxima are shifted to the side of high temperatures (till the melting temperature) (Figs. 2b, 3b), while for $SiO_2$ particles the maxima are observed at a much lower temperature, which with increasing radius (Fig. 4b). It is worth noting that with allowance for high-order polarization and correlation effects we will obtain an additional increase of radiant emittance.

Figures 2(c)—4(c) show the dependences of $\dot{Q}^{(2)}/\dot{Q}$ on the particle radius $r$ at $R = 2r$ (close contact of dimer particles) corresponding to different temperature, while Figs. 2(d)—4(d) show the dependences of $\dot{Q}^{(2)}/\dot{Q}$ on the ratio $R/r$ corresponding to different combinations of the particle radius $r$ and temperature $T$. As we can see, the presence of a neighboring particle significantly affects the RHT, since it may either partly diminish thermal radiation in vacuum ("screening effect", $\dot{Q}^{(2)}/\dot{Q} < 0$), or create a "mirror effect" ($\dot{Q}^{(2)}/\dot{Q} > 0$), when RHT increases. The sign of $\dot{Q}^{(2)}/\dot{Q}$ depends on the particle radius and temperature, while the magnitude of $\dot{Q}^{(2)}/\dot{Q}$ by modulus decreases with increasing distance $R$. A similar phenomenon is known in calculations of the optical transparency of small water droplets in a cloud [18], when the cloud may become transparent or opaque for optical radiation depending on the distance between the droplets.

In an ensemble of $N$ particles, to a first approximation, the "screening" ("mirror") effect of thermal radiation is proportional to the number of nearest neighbors $N_0$ (Eq. (7)) and $b_N/N \to 0$ for $SiO_2$ particles at $N_0 = 12$ at $r = 2\,\mu m$. I.e. we may expect that the ensemble of closely packed silica particles of this radius must be a very good heat protector. On the other hand, an ensemble of graphite particles with a radius of $0.2\,\mu m$ will be a very good heat radiator, since in this case blackness factor $b_N$ per one particle exceeds unity.

Moreover, it is worth noting that for an ensemble of $N$ nanoparticles, their total RHT in vacuum rises proportionally to $N$ (Eq. (7)), while for a solid body with the same outer radius the black body radiation rises as $N^{2/3}$. Therefore, at $N \gg 1$ the ensemble of nanoparticles should be a much more intense source (absorber) of thermal radiation than an equivalent solid body of the same dimension, while the total mass of the system of nanoparticles can be considerably lower than the mass of macroscopic body. This opens a wide perspective in tailoring such systems.

## Conclusions

We have calculated radiant emittance of dimers formed by gold, graphite and $SiO_2$ particles placed in vacuum assuming that the temperature of particles differs from that of vacuum. The presence of neighboring particles of the same temperature causes an oscillating character of radiant emittance (per one particle) depending on the particle size, interparticle distance and temperature. The effective blackness factor of particles can be higher ("mirror effect") or lower ("screening effect") than in the case of an isolated particle in vacuum. The same conclusion holds the case of an ensemble of particles. The maximum radiant emittance for dimers formed by solid gold and graphite particles (per one particle) is expected at high temperatures close to the melting points involved ( 0.92 and 0.084, respectively), and a relatively small particle radiusof 0.1 to 0.3 $\mu m$. Surprisingly, for a dimer formed by $SiO_2$ particles, a maximum blackness factor 0.55 corresponds to the temperature of about $220K$ and radius of about $3\mu m$. We conclude that an ensemble of particles could be a much more intense source of thermal radiation than an equivalent solid body with the same outer surface area. Alternatively, when the neighboring particles create a significant "screening effect" ($SiO_2$), an ensemble of particles could be a very good heat protector. This opens wide perspectives for designing heat-emitting and heat-protecting systems based on the properties of nanoparticles and their ensembles.

# FIGURE CAPTIONS

Fig.1 A scheme of particle configuration.

Fig. 2a,b Dependence of blackness factor $b$ on the particle radius $r$ at different temperature ($a$) and on the particle temperature $T$ at different radius ($b$) in the case $R = 2r$ (closest contact of particles in dimer) for $Au$ particles:

 ($a$) thin solid line-- $T = 300K$, dash thin line-- $T = 500K$; dash-dot line-- $T = 750K$, dash line-- $1000K$, solid bold line-- $T = 1337K$ (melting temperature);

 ($b$) dash-dot line-- $r = 0.1 \mu m$, dash bold line-- $r = 0.2 \mu m$, dash line-- $r = 0.3 \mu m$, solid bold line-- $r = 0.35 \mu m$, solid thin line-- $r = 0.6 \mu m$, dash thin line-- $r = 1 \mu m$.

Fig.2c,d Dependence of $\dot{Q}^{(2)}/\dot{Q}$ on the particle radius $r$ ($c$) (at $R = 2r$) and on the relative distance $R/r$ ($d$):

($c$) solid thin line-- $T = 150K$, dash thin line-- $T = 200K$, solid bold line-- $T = 250K$, dash bold line-- $T = 500K$, dash-dot line-- $T = 1000K$.

($d$) dash thin line-- $r = 0.2 \mu m, T = 500K$, solid thin line-- $r = 0.2 \mu m, T = 1300K$, dash bold line-- $r = 0.35 \mu m, T = 900K$, dash-dot line-- $r = 1 \mu m, T = 250K$, solid bold-- $r = 1.86 \mu m, T = 250K$.

Fig.3 $a,b$ Same as on Fig. 2a,b for graphite particles:

 ($a$) dot line-- $T = 200K$, dash thin line-- $T = 300K$; solid thin line-- $T = 500K$, dash-dot line-- $1000K$, dash bold line-- $T = 2000K$, solid bold line-- $T = 3000K$;

 ($b$) dash-dot line $r = 0.1 \mu m$, dash bold line-- $r = 0.15 \mu m$, solid bold line-- $r = 0.2 \mu m$, solid thin line $r = 0.25 \mu m$, dash thin line-- $r = 0.5 \mu m$, dash-dot thin line-- $r = 1 \mu m$.

Fig. 3c,d Same as on Fig. 2c,d for graphite particles

($c$) solid thin line-- $T = 200K$, solid dash line-- $T = 300K$, solid bold line-- $T = 500K$, dot line-- $T = 1000K$, dash thin line-- $T = 2000K$, dash-dot solid line-- $T = 3000K$

($d$) solid thin line— $r = 0.08 \mu m, T = 1600K$, solid bold line-- $r = 0.2 \mu m, T = 3000K$, dash line-- $r = 1 \mu m, T = 500K$, dash thin line-- $r = 1 \mu m, T = 1000K$, dash-dot line-- $r = 0.13 \mu m, T = 1000K$

Fig. 4 $a,b$ Same as on 2a,b for $SiO_2$ particles:

 ($a$) solid line 1-- $T = 223K$, solid line 2-- $T = 500K$, dash line 1-- $T = 123K$, dash line 2-- $T = 173K$; dash-dot line-- $T = 273K$; dot line 1-- $T = 323K$; dot line 2-- $T = 383K$.

 ($b$) lines 1 to 5 correspond to: 1 -- $r = 0.25 \mu m$; 2-- $r = 0.5 \mu m$; 3-- $r = 1 \mu m$; 4-- $r = 2 \mu m$; 5-- $r = 3 \mu m$

Fig. 4 *c,d* Same as on 2*c,d* for $SiO_2$ particles:

(*c*) solid thin line--$T = 173K$, dash bold line--$T = 223K$, dash-dot line--$T = 273K$, solid bold line--$T = 500K$, dash thin line--$T = 800K$;

(*d*) solid thin line--$r = 0.25 \mu m, T = 223K$, dot line--$r = 0.5 \mu m, T = 223K$, dash-dot line--$r = 2\mu m, T = 380K$, dot line--$r = 1\mu m, T = 223K$, solid bold line--$r = 1\mu m, T = 600K$

FIGURE 1

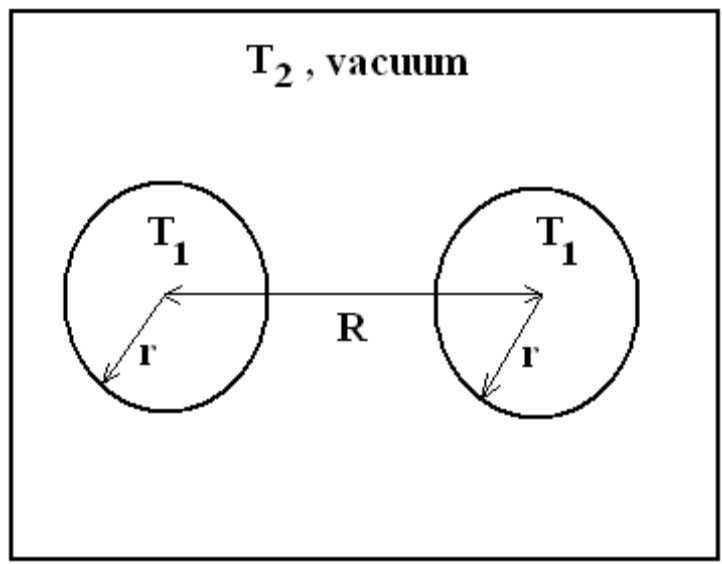

FIGURE 2(a,b)

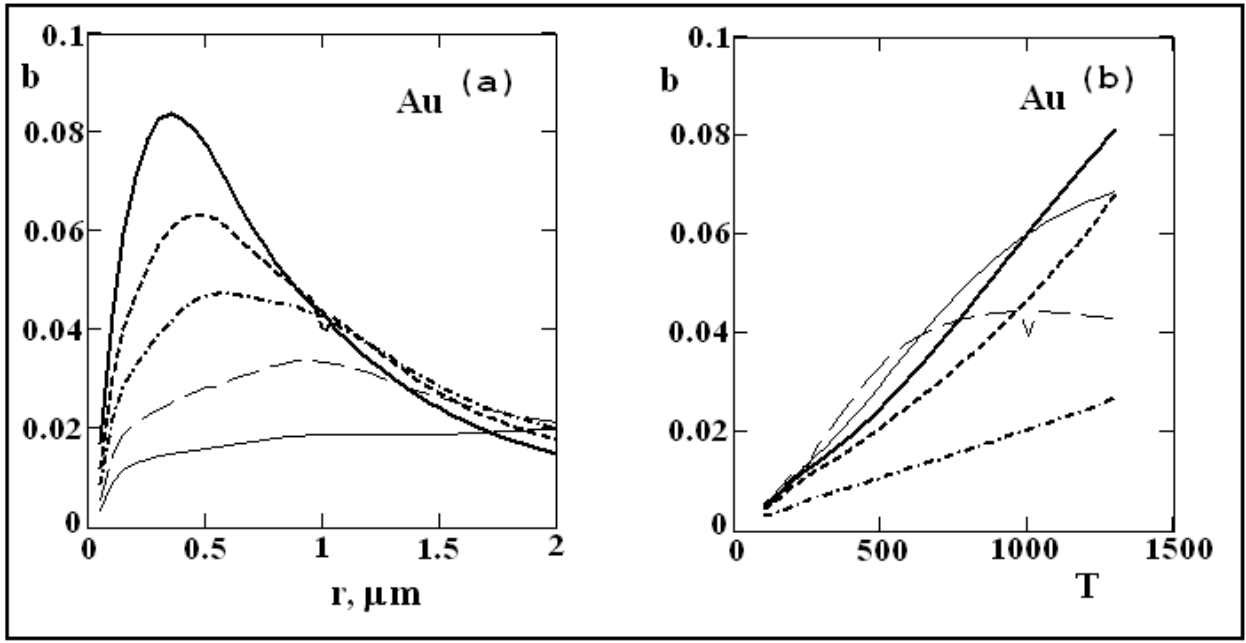

FIGURE 2(c,d)

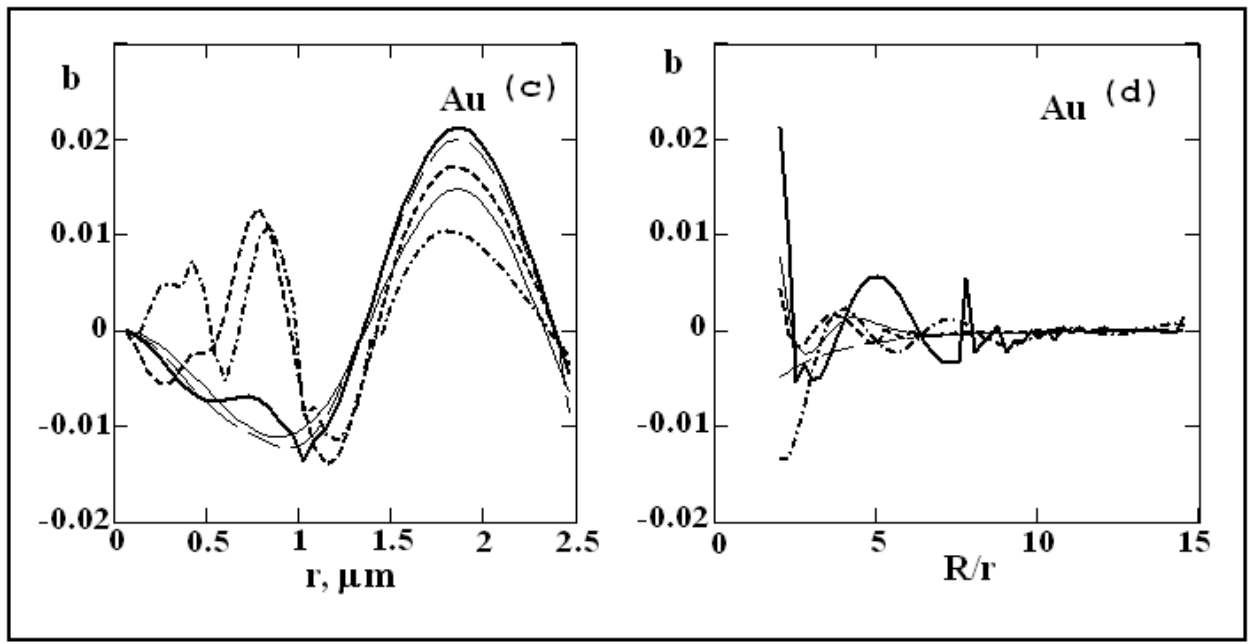

FIGURE 3(a,b)

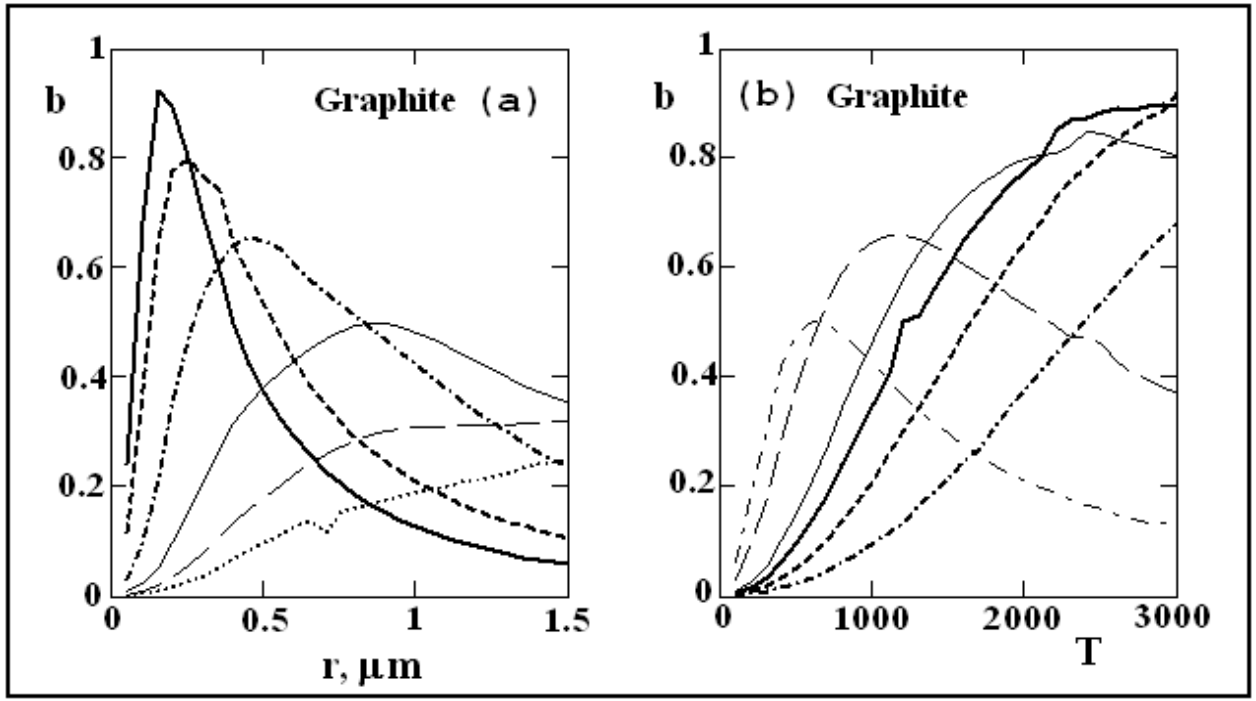

FIGURE 3(c,d)

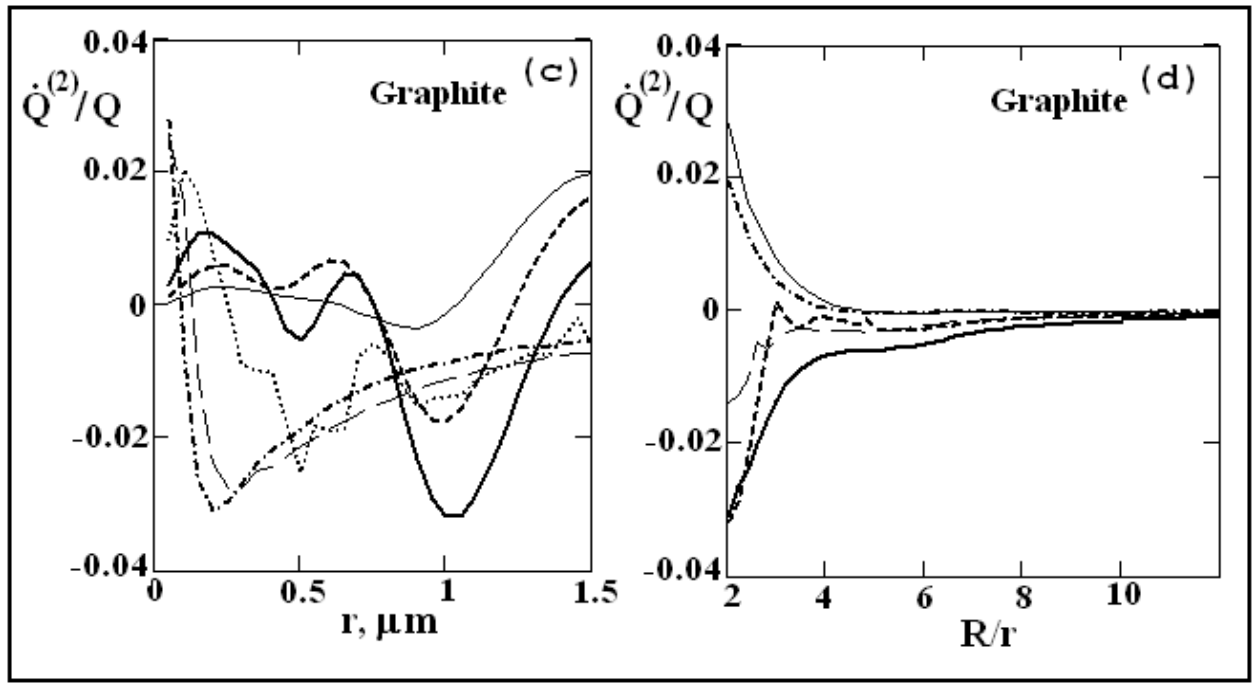

FIGURE 4(a,b)

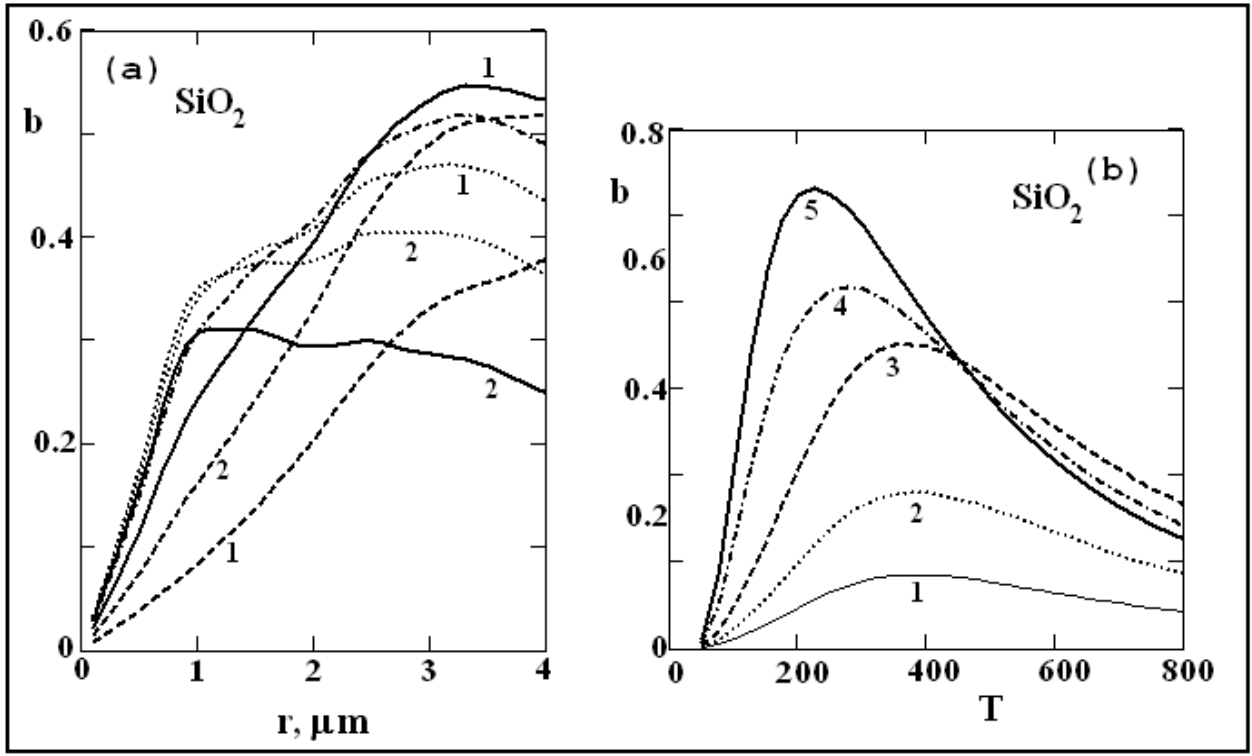

FIGURE 4(c,d)

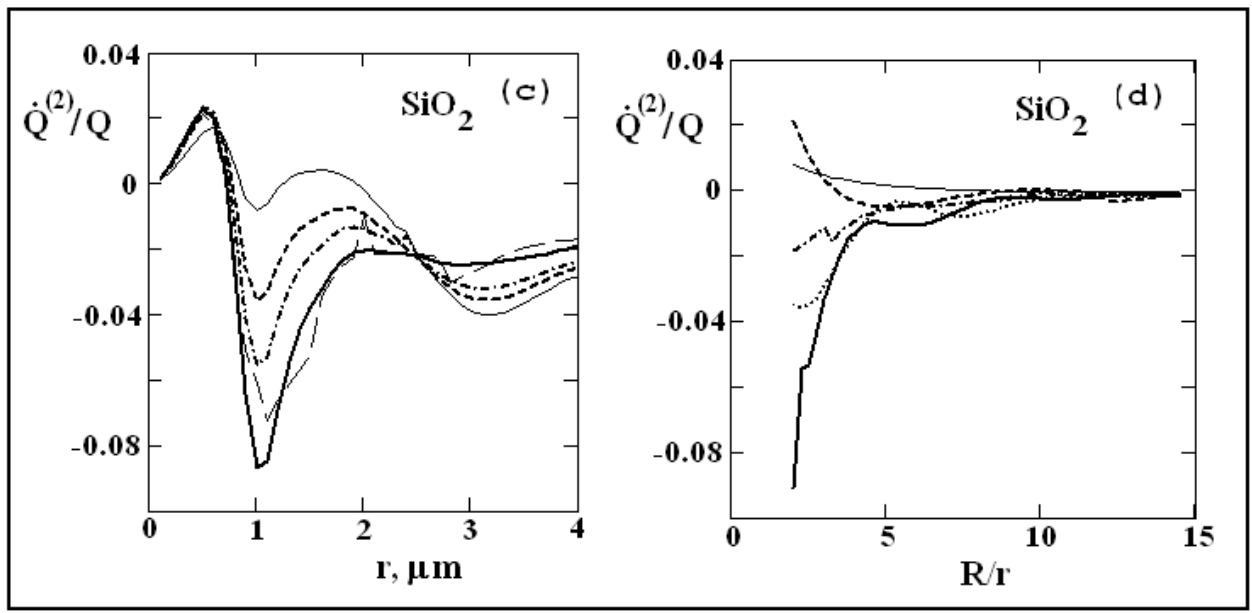